\documentclass{moriond}

\def\be{\begin{equation}}
\def\ee{\end{equation}}
\def\bea{\begin{eqnarray}}
\def\eea{\end{eqnarray}}

\newcommand{\nc}{\newcommand}
\nc{\beq}{\begin{equation}}
\nc{\eeq}{\end{equation}}
\nc{\n}{\nonumber \\}
\nc{\cm}{21$\,$cm }
\nc{\ihMpc}{h{\rm\;Mpc^{-1}}}
\nc{\Df}{\Delta f}
\nc{\Tsky}{\bar{T}_{\rm sky}}
\nc{\Ta}{\bar{T}_a}
\nc{\Tsig}{\bar{T}_{\rm sig}}
\nc{\tint}{t_{\rm int}}
\nc{\Poh}{P_{\rm HI}}
\nc{\OHI}{\Omega_{\rm HI}}
\nc{\VR}{V_{R}}
\nc{\Nfeed}{N_{\rm feed}}
\nc{\Nyear}{N_{\rm year}}
\nc{\Asur}{A_{\rm survey}}
\nc{\Lcyl}{L_{\rm cyl}}
\nc{\Wcyl}{W_{\rm cyl}}
\nc{\Apix}{A_{\rm pixel}}
\nc{\knyq}{k_{\rm Nyq}}
\nc{\hdel}{\hat{\delta}}

\newcommand{\Photo}{}

\begin{document}
\vspace*{4cm}
\title{Measuring the \cm Global Brightness Temperature Spectrum During the Dark Ages with the SCI-HI Experiment}

\author{Jeffrey B. Peterson\textsuperscript{1}, Tabitha C. Voytek\textsuperscript{1}, Aravind Natarajan\textsuperscript{1,2}, Jos\'{e} Miguel J\'{a}uregui Garc\'{i}a\textsuperscript{3}, Omar L\'{o}pez-Cruz\textsuperscript{3}}

\address{\textsuperscript{1}McWilliams Center for Cosmology, Carnegie Mellon University, Department of Physics, 5000 Forbes Ave., Pittsburgh PA 15213, USA \\ \textsuperscript{2}Department of Physics and Astronomy \& Pittsburgh Particle physics, Astrophysics and Cosmology Center, University of Pittsburgh, 100 Allen Hall, 3941 O'Hara Street, Pittsburgh, PA 15260 \\ \textsuperscript{3}Instituto Nacional de Astrof\'{i}sica, Optica y Electr\'{o}nica (INAOE), Coordinaci\'{o}n de Astrof\'{i}sica, Luis Enrique Erro No. 1 Sta. Ma. Tonantzintla, Puebla, 72840 Mexico}

\maketitle\abstract{
We present an update on the SCI-HI experiment, which is designed to measure the all-sky (global) \cm brightness temperature during the end of the Dark Ages. Results from preliminary observations in June 2013 are discussed, along with system improvements and planned future work.}

\section{Introduction}
Very few observations constrain models of the Dark Ages, during which the simple initial conditions seen with the Cosmic Microwave Background (CMB) evolved into the complex structures we see today. A key process during the Dark Ages is the formation of the first (Pop III.1) stars in dark matter minihalos of mass $\approx 10^6 - 10^8 M_\odot$ at redshift $z \approx$ 20-30. \cite{Omukai_Palla2} \cite{Bromm} These first, short-lived stars provided the Universe with the heavy elements necessary for subsequent generations of stars and planets. 

Besides providing heavy elements, the first stars impacted the intergalactic medium (IGM), transforming it from neutral hydrogen into the ionized state we see today. We can learn about star formation by examining the all-sky (global) \cm brightness temperature ($T_{\rm b}$). This is possible because the evolution of the IGM prior to reionization is characterized by decrement followed by increment in $T_{\rm b}$, described in the following process. Lyman-$\alpha$ photons, produced by the first stars, couple the spin ($T_{\rm s}$) and kinetic ($T_{\rm k}$) temperature of the IGM through the Wouthuysen-Field mechanism. \cite{lya1} \cite{lya2} Because $T_{\rm k} \propto$ $(1+z)^2$ is well below the CMB temperature ($T_{\rm \gamma}$) at $z \approx$ 20-30 this coupling causes a decrement in $T_{\rm b} \propto$ $1-T_{\rm \gamma}/T_{\rm k}$. Later the IGM is heated by x-rays or $\gamma$-rays and $T_{\rm s}$ rises above $T_{\rm \gamma}$, ending the $T_{\rm b}$ trough. Observations of the $T_{\rm b}$ trough can be made by measurement of the \cm global spectrum. \cite{shaver} \cite{pritchard_loeb_review} The predicted depth ($\approx$0-300 mK), width ($\approx$0-50 MHz), and center frequency ($\approx$60-100 MHz) are all dependent on the model of first star formation.

Several experiments seek to measure the $T_{\rm b}$ trough in the \cm global spectrum. These include the Large Aperture Experiment to Detect the Dark Ages (LEDA), \cite{leda} an expanded EDGES, \cite{edges} the LOFAR Cosmic Dawn Search (LOCOS), \cite{lofar} \cite{vedantham} ``Sonda Cosmol\'{o}gica de las Islas para la Detecci\'{o}n de Hidr\'{o}geno Neutro'' (SCI-HI), \cite{sci-hi} and the proposed Dark Ages Radio Explorer (DARE). \cite{dare} In the following sections, we will focus on the SCI-HI global spectrum experiment. We will discuss the recent results from the first experiment deployment, \cite{sci-hi} the experiment structure, and describe recent system improvements. 

\section{Experimental Setup}
\begin{figure}
\centering
\begin{minipage}[b]{0.47\textwidth}
\centering
\includegraphics[width=0.75\linewidth]{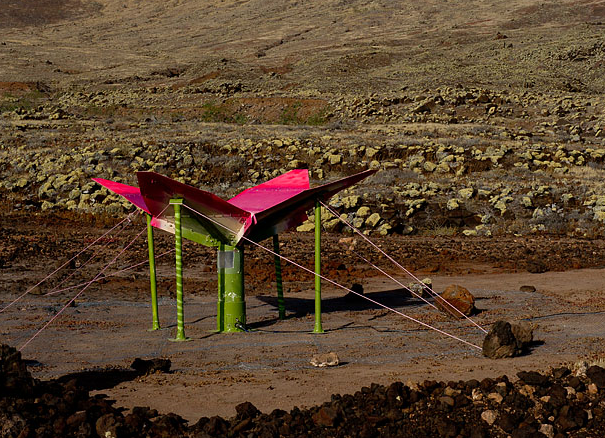}
\caption{HIbiscus Antenna on-site on Isla Guadalupe, Mexico.}
\label{antenna}
\end{minipage}%
\begin{minipage}[b]{0.02\textwidth}
\hspace{1cm}
\end{minipage}%
\begin{minipage}[b]{0.47\textwidth}
\centering
\includegraphics[width=0.75\linewidth]{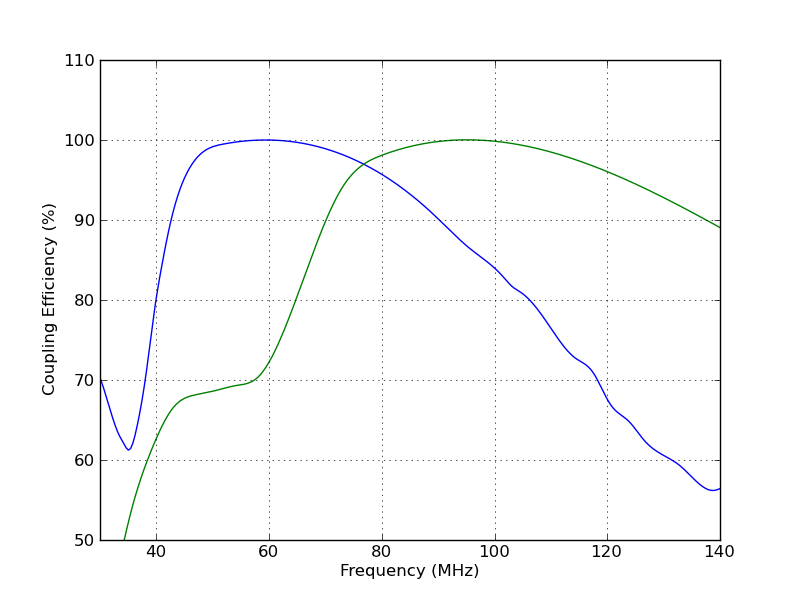}
\caption{Antenna coupling efficiencies ($\eta (\nu)$) for two scaled HIbiscus antennas, calculated using the measured antenna and amplifier impedances.}
\label{efficiency}
\end{minipage}
\end{figure}

In order to probe the \cm global spectrum, we designed a single antenna experiment. Because the HI global spectrum structure is distributed over a wide frequency range, it was necessary to develop a specialized antenna. Several antenna designs were considered, modelled and tested in the lab and field; log periodic, fat inverted vee, horn, sleeve dipole, and four square antennas. Our best results were obtained by modifying a four-square design into the HIbiscus antenna (see Fig. \ref{antenna}). Working with a finite element simulation software and scale model testing, the HIbiscus antenna was tuned to provide an optimal combination of bandwidth, impedence and beam shape. This design has a smooth $55^{\circ}$ beam (FWHM) at its center frequency and a coupling efficiency ($\eta(\nu)$) above 90\% for nearly an octave of bandwidth. 

The SCI-HI experiment is intended to collect data from 40-130 MHz, larger than the bandwidth of a single HIbiscus antenna. To solve this problem, we built scaled copies of the HIbiscus design with center frequencies at 70 MHz and 100 MHz, allowing data collection over the entire band. The antennas have $\eta(\nu)$ above 90\% for 55-90 MHz and 70-140 MHz respectively, as is shown in Fig. \ref{efficiency}. 

Besides allowing data collection over a wider band, the use of overlapping antennas allows a cross-check of data. Signals from the sky will appear in datasets from both antennas in the overlap region (70-90 MHz). A single sampling system is used alternately with either of the scaled antennas. This limits the rate of data collection but provides consistency between the two datasets. Additional details of the sampling system can be found in Voytek et al. \cite{sci-hi} 

\section{Deployment Sites}
\begin{figure}
\centering
\begin{minipage}[b]{0.47\linewidth}
\centering
\includegraphics[width=0.95\linewidth]{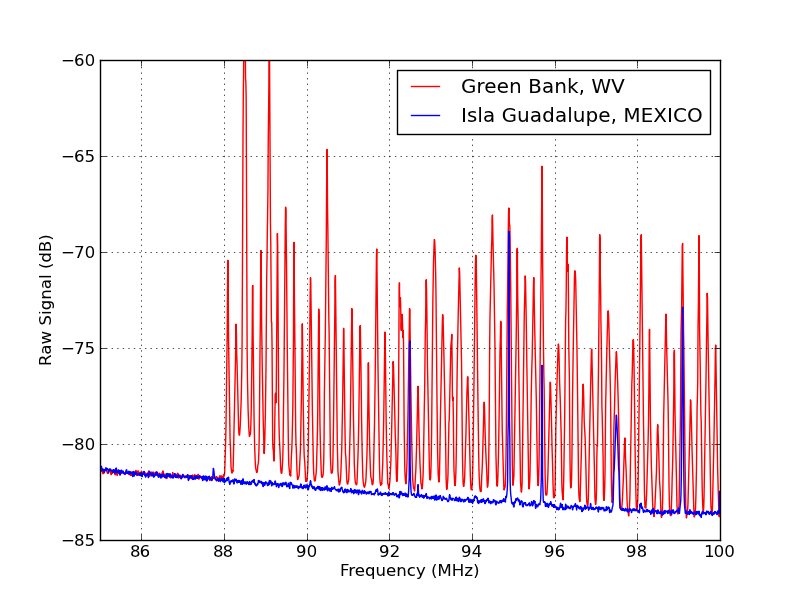}
\caption{Comparison of RFI in the FM frequency band between Green Bank, WV and Isla Guadalupe, Mexico. Except for a few stations the noise on Guadalupe is about 0.1 dB above foregrounds, while Green Bank is $\sim$10 dB above foregrounds.}
\label{RFI}
\end{minipage}%
\begin{minipage}[b]{0.02\textwidth}
\hspace{1cm}
\end{minipage}%
\begin{minipage}[b]{0.47\linewidth}
\centering
\includegraphics[width=0.75\linewidth]{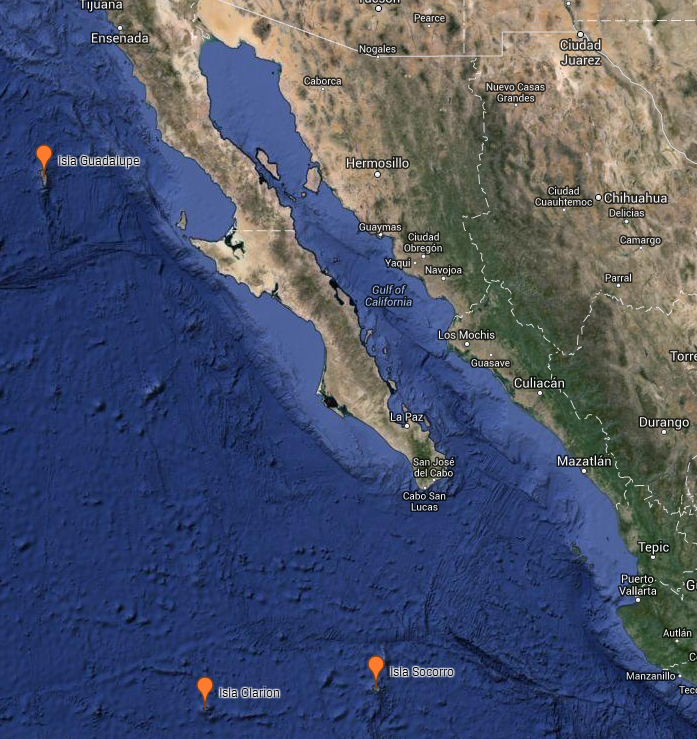}
\caption{Map of eastern Pacific Ocean, including Isla Guadalupe, Isla Socorro and Isla Clari\'{o}n.}
\label{map}
\end{minipage}
\end{figure}

In the frequency range of 40-130 MHz, major interfering flux contributions include broadcast emissions from television stations and FM radio stations which transmit in our band. Even at the U.S. National Radio Quiet Zone in Green Bank, West Virginia, the FM signal exceeds the sky signal by 10 dB over the entire FM band of 88-108 MHz (see Fig. \ref{RFI}). The equivalent $T_{\rm b}$ averages about 20,000 Kelvin. To achieve an RFI level below the \cm signal requires deployment to a significantly isolated radio-quiet location. We achieve this isolation by travelling to lightly inhabited islands located far from the mainland. Our islands of choice are located in the eastern Pacific Ocean, off the coast of Mexico (see Fig. \ref{map}). 

We have already deployed twice to Isla Gua\-da\-lu\-pe, the northern-most of our potential sites. Isla Gua\-da\-lu\-pe (Latitude $28^{\circ}$ $58'$ $24''$ N, Longitude $118^{\circ}$ $18'$ $4''$ W), is 260 km off the Baja California peninsula in the Pacific Ocean. It is a Mexican biosphere reserve and has a population of less than 100; ecology-resoration teams, Mexican Marines, and members of a fishing cooperative. We spent two weeks on Isla Gua\-da\-lu\-pe (June 1-15, 2013) observing with SCI-HI on the western side of the island. Even at this remote site, we still detect some RFI from the mainland; although residual FM is only about 0.1 dB ($\leq$70 K) above the Galactic foreground level (see Fig. \ref{RFI}).

Additional distance from the Mexican coast can provide further attenuation of RFI. Isla Socorro (Latitude $18^{\circ}$ $48'$ $0''$ N, Longitude $110^{\circ}$ $90'$ $0''$ W), is $\sim$600 km off the Pacific coast of Mexico. This island is also a biosphere reserve and has similarly minimal infrastructure to Isla Guadalupe. Operation of SCI-HI at this location should provide an RFI environment sufficiently quiet in our frequency band. 

Isla Socorro may still not be quiet enough due to insufficient mainland distance or RFI from the island's Mexican naval contingent. If this proves to be the case, Isla Clari\'{o}n (Latitude $18^{\circ}$ $22'$ $0''$ N, Longitude $114^{\circ}$ $44'$ $0''$ W) is $\sim$700 km off the coast, has a much smaller naval contingent, and can be utilized as well.

\section{Current Results}
\begin{figure}
\centering
\begin{minipage}[b]{0.47\linewidth}
\centering
\includegraphics[width=0.9\linewidth]{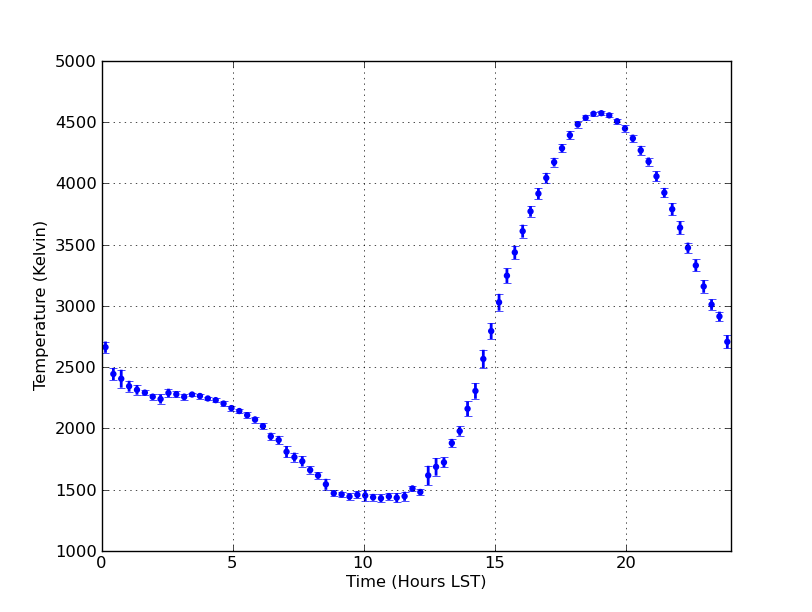}
\caption{Diurnal variation of a single 2 MHz wide bin centered at 70 MHz. Calibrated mean with RMS error bars from day-to-day variation are shown for 9 days of observation binned in $\sim$ 18 minute intervals. Larger error bars correspond to LSTs where the quantity of useable data is smaller.}
\label{diurnal}
\end{minipage}%
\begin{minipage}[b]{0.02\textwidth}
\hspace{1cm}
\end{minipage}%
\begin{minipage}[b]{0.47\linewidth}
\centering
\includegraphics[width=0.9\linewidth]{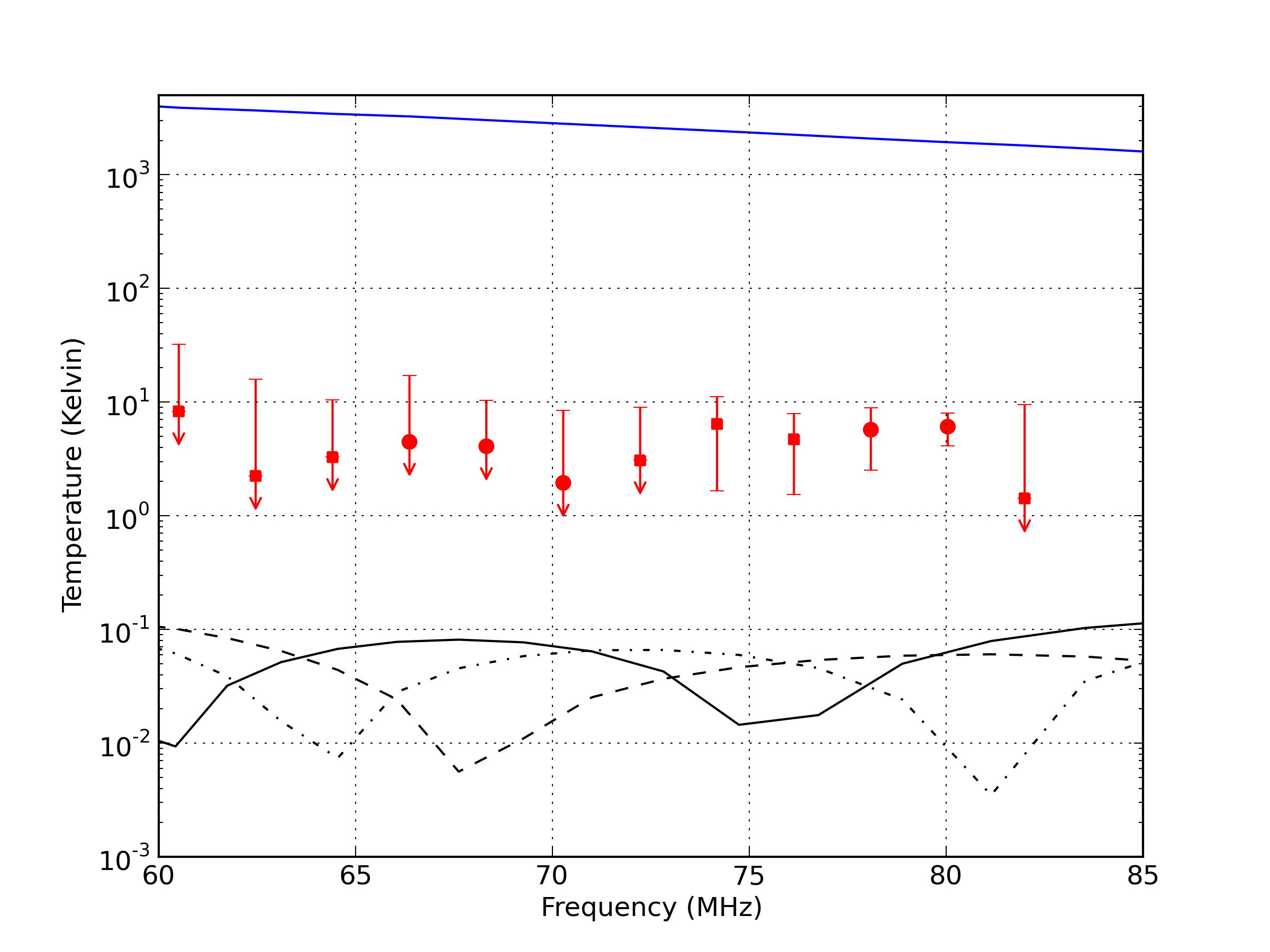}
\caption{Log magnitude comparison of foregrounds (blue), residuals from 4.4 hours of integration (red) and predictions from three different reionization models (black). For the residuals, circles are positive values and squares are negative values and error bars show the daily variance.}
\label{comb_resid}
\end{minipage}
\end{figure}

Data is processed through a series of steps. First, the data is cleaned to excise RFI contaminated data. Second, amplifier noise is removed and the data is corrected using the coupling efficiency ($\eta (\nu)$). Third, brightness response is calibrated using the Galactic Global Sky Model (GSM) \cite{gsm} and the simulated antenna beam pattern.  This calibration is done using the diurnal variation of the data (see Fig. \ref{diurnal}), and is analagous to an on-source/off-source calibration of a single dish radio telescope. 

Data processing is done independently for each day of data. After calibration, fitting the daily mean spectrum to a log-polynomial shape with three terms yields residuals that are dominated by systematic uncertainty. Residual mean and variance, weighted for exposure, are shown in Fig. \ref{comb_resid}, along with the foreground signal prior to fit. Predictions from three cosmological models with mean brightness temperature removed, obtained using the {\scriptsize SIMFAST} code, \cite{simfast} \cite{21cmfast} are also shown. 

To check that data processing does not cause significant attenuation to the \cm signal, a simulation calibration technique \cite{gmrt3} was used. This technique adds a simulated signal of appropriate shape (and varying magnitude) to the data prior to processing and compares residuals before/after addition of the simulated signal. Results from this check showed that $\approx$75\% of the added signal remains in the residuals. For more detail on data processing, see Voytek etal. \cite{sci-hi}

\section{Discussion and Future Work}
The first results from the SCI-HI experiment yield residuals of $\approx$5-10 Kelvin with 4.4 hours of integration. Given that the foregrounds have brightness temperatures $>$1000 Kelvin, this means that residuals are $<$1\% of foreground signal. Residual levels are lower than early predictions based on possible spatial structure in the spectral index of the foregrounds. \cite{Liu_etal} This is a promising sign for the future of \cm global experiments. 

Despite the promise of the first results from the SCI-HI experiment, a further decrease in residuals of 1-2 orders of magnitude is necessary to measure the \cm signal (300 mK peak-to-peak temperature amplitude). As calculated using the radiometer equation, thermal noise is $\sim$2 orders of magnitude below the current residual levels. This indicates that residuals are dominated by systematic errors and residual foregrounds. Improvements to the apparatus are underway, and should help decrease systematic errors. 

Systematic errors can be described in two major categories. First, the experiment faraday cage does not contain self-generated RFI to sufficient levels. In particular, self-generated RFI contributes substantially to the residuals above 90 MHz and can also be found minimally at lower frequencies. This self-generated RFI varies with time and introduces frequency dependent structures. Improvements are therefore in progress to the sampling system and its shielding. Repeated deployment to quiet sites is necessary to test the system because the low level of the self-generated RFI makes it undetectable in lab tests, where the noise is dominated by strong broadcast signals.

Second, much of the residuals are related to calibration error associated with the lack of full 24 hour days of data. During this preliminary data collection, the system power source was 12 V lead-acid batteries, which yielded time variable voltages and caused frequent gaps in data collection. Improvements to the sampling system and replacement of our power generation system with a stable power source will facilitate the collection of complete days of data, thereby reducing the calibration errors. 

Deployment of the SCI-HI experiment is planned for Isla Socorro and Isla Clari\'{o}n, taking advantage of the additional distance from the mainland. These islands are expected to provide an RFI environment where the FM band signals are below the thermal noise level, allowing a substantial increase in the frequency range studied. With two full weeks of data and complementary datasets from multiple HIbiscus antennas, our goal is to achieve residuals below the 100 mK level from 40-130 MHz.

\section*{Acknowledgements}
Travel to Isla Guadalupe would not have been possible without support from local agencies in Mexico, including Grupo de Ecolog\'{i}a y Conservaci\'{o}n de Islas A.C. (GECI), Secretar\'{i}a de Marina (SEMAR), Secretar\'{i}a de Gobernaci\'{o}n (SEGOB), Comisi\'{o}n Nacional de Areas Naturales Protegidas (CONANP), Reserva de la Biosfera de la Isla Guadalupe, Sociedad Cooperativa de Producci\'{o}n Pesquera de Participaci\'{o}n Estatal Abuloneros y Langosteros, S.C.L., and Dr. Ra\'{u}l Michel. 

A.N., J.B.P., and T.C.V. acknowledge funding from NSF grant AST-1009615. A.N. thanks the McWilliams Center for Cosmology for partial financial support. J.M.G.C. and O.L.-C. thank INAOE for financial support. J.M.G.C. acknowleges a CONACyT Beca Mixta that allowed him to spend a year at the McWilliams Center for Cosmology

\section*{References}
\bibliography{references}

\end{document}